\documentclass[11.5pt,twocolumn]{quantumarticle}
\pdfoutput=1 
\usepackage[T1]{fontenc}

\usepackage{tikz}
\usetikzlibrary{positioning, calc}
\usetikzlibrary{calc}
\usetikzlibrary{arrows}
\usepackage{tikz-3dplot}
\usepackage{graphicx}
\usepackage{mdwlist}
\usepackage{color}
\usepackage{thmtools}
\usepackage{amssymb}
\usepackage{physics}
\usepackage{amsmath}
\usepackage{array}
\usepackage{url,hyperref}
\usepackage[numbers,sort&compress]{natbib}
\usetikzlibrary{fadings}
\usetikzlibrary{decorations.pathmorphing}
\usepackage{mathrsfs}
\usepackage{authblk}
\usepackage{amsmath}
\usepackage{amsthm}
\usepackage{mathtools}
\usepackage{enumitem}

\usepackage{cleveref}

\DeclareMathOperator{\poly}{poly}
\DeclareMathOperator{\polylog}{polylog}
\DeclareMathOperator{\fm}{fam}
\DeclareMathOperator{\anc}{anc}

\DeclareMathOperator{\pr}{par}
\DeclareMathOperator{\crc}{circ}

\DeclareMathOperator{\supp}{supp}

\usetikzlibrary{decorations.pathreplacing}
\tikzset{snake it/.style={decorate, decoration=snake}}

\usetikzlibrary{decorations.pathreplacing,decorations.markings}

\tikzset{
    >=stealth',
    punkt/.style={
           rectangle,
           rounded corners,
           draw=black, very thick,
           text width=6.5em,
           minimum height=2em,
           text centered},
    pil/.style={
           ->,
           thick,
           shorten <=2pt,
           shorten >=2pt,},
  on each segment/.style={
    decorate,
    decoration={
      show path construction,
      moveto code={},
      lineto code={
        \path [#1]
        (\tikzinputsegmentfirst) -- (\tikzinputsegmentlast);
      },
      curveto code={
        \path [#1] (\tikzinputsegmentfirst)
        .. controls
        (\tikzinputsegmentsupporta) and (\tikzinputsegmentsupportb)
        ..
        (\tikzinputsegmentlast);
      },
      closepath code={
        \path [#1]
        (\tikzinputsegmentfirst) -- (\tikzinputsegmentlast);
      },
    },
  },
  mid arrow/.style={postaction={decorate,decoration={
        markings,
        mark=at position .5 with {\arrow[#1]{stealth'}}
      }}}
}

\usepackage{hyperref}
\usepackage{slashed}

\usepackage{float}		
\usepackage{graphicx}		
\usepackage{subcaption}
\usepackage{amsmath}

\newcommand{\defi}{:=}
\usepackage{bbold}

\newtheorem{theorem}{Theorem}

\newtheorem{definition}[theorem]{Definition}

\newtheorem{primitive}[theorem]{Primitive}

\theoremstyle{definition}
\newtheorem{protocol}[theorem]{Protocol}

\begin{document}

\title{Non-local computation of quantum circuits with small light cones}

\author[1]{Kfir Dolev}
\orcid{0000-0002-4030-5410}
\author[1]{Sam Cree}
\orcid{0000-0003-2283-3903}

\affil[1]{Stanford University}

\begin{abstract}
The task of non-local quantum computation requires implementation of a unitary on $n$ qubits between two parties with only one round of communication, ideally with minimal pre-shared entanglement.
We introduce a new protocol that makes use of the fact that port-based teleportation costs much less entanglement when done only on a small number of qubits at a time.
Whereas previous protocols have entanglement cost independent of the unitary or scaling with its complexity, the cost of the new protocol scales with the \emph{non-locality} of the unitary.
Specifically, it takes the form $\sim n^{4V}$ with $V$ the maximum volume of a past light cone in a circuit implementing the unitary.
Thus we can implement unitary circuits with $V\sim O(1)$ using polynomial entanglement, and those with $V\sim \polylog(n)$ using quasi-polynomial entanglement.
For a general unitary circuit with $d$ layers of $k$-qubit gates $V$ is at most $k^d$, but if geometric locality is imposed it is at most polynomial in $d$.
We give an explicit class of unitaries for which our protocol's entanglement cost scales better than any known protocol. 
We also show that several extensions can be made without significantly affecting the entanglement cost -- arbitrary local pre- and post-processing; global Clifford pre- and post-processing; and the addition of a polynomial number of auxiliary systems.
\end{abstract}

\vfill

\maketitle

\pagebreak

\tableofcontents

\section{Introduction}

Non-local quantum computation (NLQC) is a quantum task with both practical applications and relevance to fundamental physics \cite{tagging,location-dependent,PBQC,beigi-koenig,tagging2,conundrum,loss-tolerant,practical-PBQC,banach-spaces,beating-classical,new-ideas,insecurity}.
Practically, it is a fundamental primitive for position-based quantum cryptography \cite{kent2006tagging, PBQC}.
In fundamental physics, it appears to have deep connection with the holographic principle \cite{qtHolography,May-pennington-sorce}.

The NLQC task involves two individuals that are given a quantum state, and must apply to it a given quantum computation with only one round of communication.
Critically, it can be shown that for some unitaries the two individuals can only accomplish this if they share a certain amount of entanglement ahead of time \cite{qtHolography}.
Knowing exactly how much entanglement is necessary is the crux of determining whether a large class of position-based cryptography schemes is viable.

Currently, linear lower bounds \cite{qtHolography,monogamy-of-entanglement} and exponential upper bounds \cite{beigi-koenig} are known for the number of shared Bell pairs needed as a function of the number of qubits the computation acts on.
However, if the structure of the computation is taken into account the upper bounds may be tightened.
For example, there are protocols that need only entanglement exponential in the T-depth or T-count of the unitary \cite{T-gate-protocol}, and others that can efficiently perform a certain class of unitaries related to classical complexity classes \cite{Logspace-routing} and \cite{code-routing}.
Furthermore, arguments from high energy physics are suggestive that the entanglement cost upper bound should be polynomial in the circuit depth for any unitary \cite{holo-nlqc}.

In this work we introduce a new protocol capable of efficiently performing NLQC for a new class of unitaries.
Our protocol is a generalization of the one introduced in Ref.~\cite{beigi-koenig}, which relies on a form of quantum teleportation known as \emph{port teleportation} \cite{port-based-teleportation}.
Unlike standard quantum teleportation, in which a teleported system is encrypted by the presence of an unknown Pauli operator, in port-based teleportation the teleported system is encrypted in that it appears in one of many systems or ``ports'', the identity of which is unknown to the receiver.
Unitaries can be easily implemented on a port-teleported system by transversally applying the unitary in question to all possible ports, whereas after standard teleportation the receiver would need to first undo an unknown Pauli encryption.
However, easier implementation of unitaries comes at the cost of requiring entanglement exponential in the number of qubits being teleported, which is the origin of the exponential cost of the protocol in Ref.~\cite{beigi-koenig}. By contrast, our protocol makes use of the circuit structure of a unitary, allowing us to port-teleport only a small batch of qubits at a time.

At a high level, our protocol consists of Alice and Bob iterating through the gates of the circuit, teleporting back and forth the batch of qubits on which each gate acts.
Alice always normal-teleports to Bob, while Bob always port-teleports to Alice.
In this way Bob always knows the identity of the correct ports, and thus can port-teleport the right batch of systems to Alice.
Meanwhile Alice always knows the identity of the Pauli encryptions, and thus can remove them to perform computations.
However, because Alice does not know the identity of the correct port, she must normal-teleport all ports back to Bob, resulting in her obtaining Pauli keys that \textit{depend on the port}, and thus she does not know which key to use when receiving the system at a later step.
We deal with this issue via the re-indexing technique of \cite{dolev}, i.e.\ by having Bob port-teleport the system through a shared resource labeled by the correct port.
Thus when Bob port-teleports a new batch of qubits to Alice, they must have set up a resource state for each combination of port values of all the port-based teleportations in the causal past of these qubits.
The number of port teleportation resource states required to do this is proportional to an exponential of the number of such teleportations, which is the number of gates in the past light cone of the current gate. Furthermore, the size of the port teleportation resource state scales exponentially with $k$, the number of qubits each gate in the circuit acts on.
This results in a total entanglement cost scaling like $n^{4V}$, with $V$ the volume of the largest past light cone in the circuit -- that is, the total number of qubits acted on by gates in the causal past of a given gate, where the same qubit can be counted multiple times.

In the worst case, the past light cone volume scales as $k^d$ with $k$ the number of qubits each gate in the circuit acts on, and $d$ the depth of the circuit.
Since the base of the exponent scales with $n$, functions implementable with only polynomial entanglement must have $O(1)$ layers of $O(1)$-sized gates.
However, the number of efficiently implementable unitaries is much larger if we mildly relax the notion of ``efficient'' to include quasi-polynomial entanglement -- that is, entanglement that scales as an exponential of a polylogarithmic function.
This class grows only mildly faster than polynomials, and still much slower than exponential functions%
\footnote{For example, $n^m$ is larger than $\exp(\log(n)^2)$ until $n=e^m$.}%
.
One class of unitaries that our protocol efficiently performs (i.e.\ using a quasi-polynomial number of Bell pairs) is constant-depth circuits of polylogarithmic sized gates.

In \cref{sec: background}, we formally introduce the problem of NLQC and the primitives of normal and port-based teleportation, and review the protocol of \cite{beigi-koenig} that ours generalizes.
We present the protocol in \cref{sec: protocol} and analyze its performance, calculating and analyzing the entanglement cost and comparing with existing protocols.
In \cref{sec:extensions}, we give several enhancements to the protocol that increase the class of unitaries it can perform efficiently: (i) arbitrary local pre- and post-processing, whereby Alice and Bob are free to act on their initial states before and after the protocol; (ii) arbitrary global Clifford pre- and post-processing; and (iii) use of a polynomial number of auxiliary registers.



\section{Background}\label{sec: background}
\subsection{Non-local quantum computation}\label{subsec: NLQC}

A non-local quantum computation (NLQC) task involves two parties, Alice and Bob.
At the start, Alice holds a quantum systems $A$ and $R_A$ while Bob holds $B$ and $R_B$.
The systems $AB$ collectively contain $n$ qubits, and for simplicity we assume each party has $n/2$ qubits.
They are also potentially entangled with a reference state $R$ that neither of them hold.
These start out in a state $\ket{\psi}_{ABR}\otimes\ket{\phi}_{R_AR_B}$.
The state $\ket{\psi}_{ABR}$ is unknown to Alice and Bob, but they are free to choose the ``resource'' state $\ket{\phi}_{R_AR_B}$.  
Alice and Bob are told of a unitary $U_{AB}$ acting on the system $AB$, which they are to perform in the following restricted manner.
First, Alice performs a quantum channel of her choice $\mathcal{N}_{AR_A\rightarrow A'M_A}$ on the systems she holds to produce an output system $A'$ that she keeps and a message system $M_A$ that she sends to Bob.
Similarly Bob performs $\mathcal{N}_{BR_B\rightarrow B'M_B}$ to obtain $B'$ and $M_B$.
Critically, the message systems do not interact during the exchange, which is why the computation is ``non-local''.
Alice then performs a channel $\mathcal{N}_{A'M_B\rightarrow A}$ and Bob the channel $\mathcal{N}_{B'M_A\rightarrow B}$.
They succeed in the task if the final state of $ABR$ is $U_{AB}\ket{\psi}_{ABR}$.
For brevity, we henceforth omit the reference state $R$ and refer to $AB$ as a pure state; however our results still apply in the more general setting.

\subsection{Teleportation primitives}

We make heavy use of two types of quantum teleportation: normal teleportation \cite{normal-teleportation} and port-based teleportation \cite{port-based-teleportation}.

\begin{primitive} Normal teleportation

\begin{enumerate}
    \item \textbf{Setup:} Alice holds an $n_A$-qubit system $A$ which, together with some reference system $R$ is in a state $\ket{\psi}_{AR}$. Alice and Bob each hold one of a maximally entangled pair of $2^{n_A}$ dimensional systems $A'B'$. 
    \item Alice performs some joint measurement on $A$ and $A'$, receiving the identity of an $n_A$-qubit Pauli operator $P$ drawn uniformly from all such operators.
    \item Bob's half of the maximally entangled state, together with the reference system,
    is then in the state $P\ket{\psi}_{B'R}$, where $P$ acts on the $B'$ subsystem.
\end{enumerate}

\end{primitive}

\begin{primitive} Port-based teleportation

\begin{enumerate}
    \item \textbf{Setup:} Alice holds an $n_A$-qubit system $A$ which, together with some reference system $R$ is in a state $\ket{\psi}_{AR}$. Alice and Bob each hold one half of $N$ maximally entangled pairs of $2^{n_A}$ dimensional systems $\{A'_xB'_x\}$. 
    \item Alice performs some joint measurement on $A$ and $\{A'_x\}$, receiving a random number $x^*\in [N]$\footnote{We use the notation $[N]$ to denote the set $\{1,\ldots,N\}.$}.
    \item Bob's half of the $x^*$th entangled system, together with the reference system, is then close to the state $\ket{\psi}_{B_{x^*}R}$, with error decreasing with $N$.
\end{enumerate}
Specifically, if $\mathcal{E}:\mathcal{B}(A)\rightarrow\mathcal{B}(B'_{x^*})$ is the CPTP map describing the error channel, then from \cite{beigi-koenig} we have
\begin{align}\label{eq: port-error}
    ||\mathcal{E}-\mathcal{I}_{\mathbb{C}^{2^{\scriptstyle{n}_{\! \text{\tiny{\(A\)}}}}}}||_\diamond\leq 4\frac{2^{2n_A}}{\sqrt{N}}.
\end{align}
We say that the system $A$ has been \emph{port-teleported}.

\end{primitive}

\subsection{Best known protocol for general unitary}\label{subsec: best-known-protocol}

The best known NLQC protocol for a generic unitary is due to Beigi and Koenig in Ref.~\cite{beigi-koenig}, and proceeds as follows.

\begin{protocol}Beigi-Koenig
\begin{enumerate}
    \item Alice normal-teleports $A$ to Bob, receiving Pauli correction $P$.
    \item Bob port-teleports the $n$-qubit system $AB$ to Alice, recieving port key $x^*$.
    \item Alice then applies $P^\dagger$ to the $A$ part of all the ports on her side. She then applies $U$ to every port.
    \item \emph{Communication Round:} Alice sends the $B$ side of every port to Bob, and Bob sends $x^*$ to Alice. Both sides then throw away all systems not coming from the $x^*$ port, completing the protocol. 
\end{enumerate}

From \cref{eq: port-error} we see that this protocol requires $N= 16 \frac{2^{4n}}{\epsilon^2}$ in order to obtain some constant error $\epsilon$.

\end{protocol}

\section{The protocol and its cost} \label{sec: protocol}

\subsection{The protocol}\label{subsec: the-protocol}
Because of the important role that circuit decompositions of a unitary plays in our protocol, we carefully define some formal terminology for use throughout.
\begin{definition}
Let a \emph{circuit decomposition} of $U$ be $\{U_j^i\}_{i\in [d],j\in [n_i]}$, meaning $U = U^d\cdots U^1$, with each $U^i$ being a product of $n_i$ gates $U^i = U^i_1 \cdots U^i_{n_i}$ each with disjoint support on $k_{i,j}$ qubits%
.
Denote the support\footnote{By support we mean the set of qubits that the operator acts on.} of $U^i_j$ as $S_j^i = \supp(U^i_j) $, the collection of which we refer to as the circuit structure $\{S_j^i\}_{i\in [d],j\in [n_i]}$.
\end{definition}
We will see that the circuit structure completely specifies the entanglement cost of our protocol.
Specifically, it depends on causal properties of the circuit structure that we characterize as follows.

\begin{figure}
    \centering
    \includegraphics[width=0.5\textwidth]{
    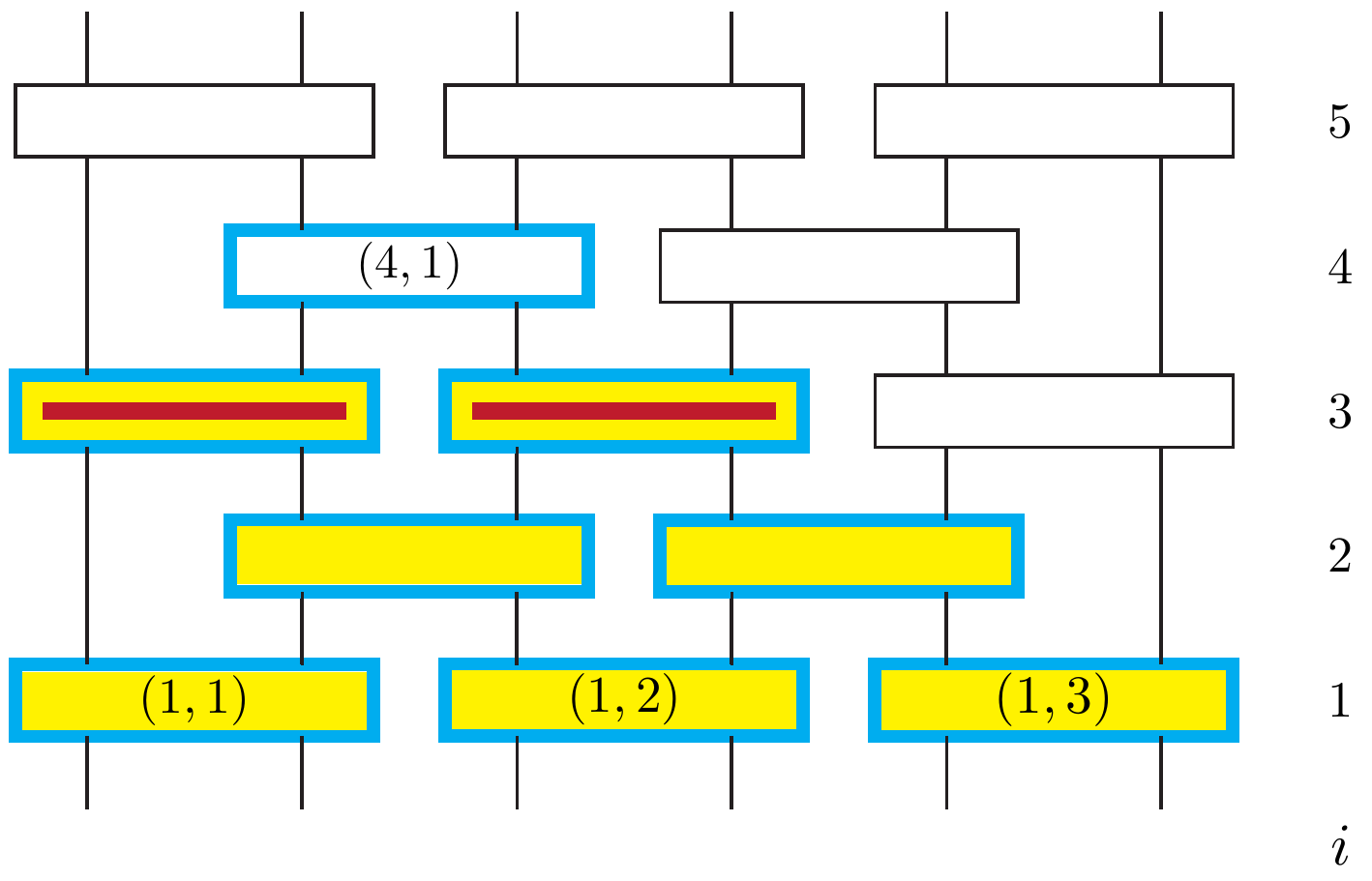}
    \caption{Various sets of interest for gate $(4,1)$. Blue: its family. Yellow: its ancestors. Red: its parents.}
    \label{fig:circuit}
\end{figure}

For low-depth circuits, a given gate acts on qubits whose state only depends on a small number of previous gates (as well as the initial state), sometimes called the past light cone.
We call the subsystems corresponding to these gates in the past light cone the \emph{ancestors} of the subsystem in question, defined formally as follows, and depicted in \cref{fig:circuit}.
\begin{definition}
For a circuit structure $\{S_j^i\}_{i\in [d],j\in [n_i]}$, we denote the \emph{parents} $\pr(i,j)$ of a subsystem $S_j^i$ at circuit layer $i$ and with subsystem label $j$, to be the set of all subsystems from the previous layer $S_{j'}^{i-1}$ with nonzero overlap, i.e. $S_j^i \cap S_{j'}^{i-1} \neq \emptyset$.
The \emph{ancestors} $\anc(i,j)$ of $S_j^i$ is then the set of all parents, parents' parents, and so on; formally, it is defined recursively as 
\begin{align*}
    \anc(i,j) = \pr(i,j) \cup  \anc(\pr(i,j) )
\end{align*}
Finally, we refer to the union of a subsystem with its ancestors as its \emph{family},  $\fm(i,j) = \anc(i,j) \cup \{ (i,j) \} $.
\end{definition}


\begin{figure*}[t!]
    \centering
    \includegraphics[width=\textwidth]{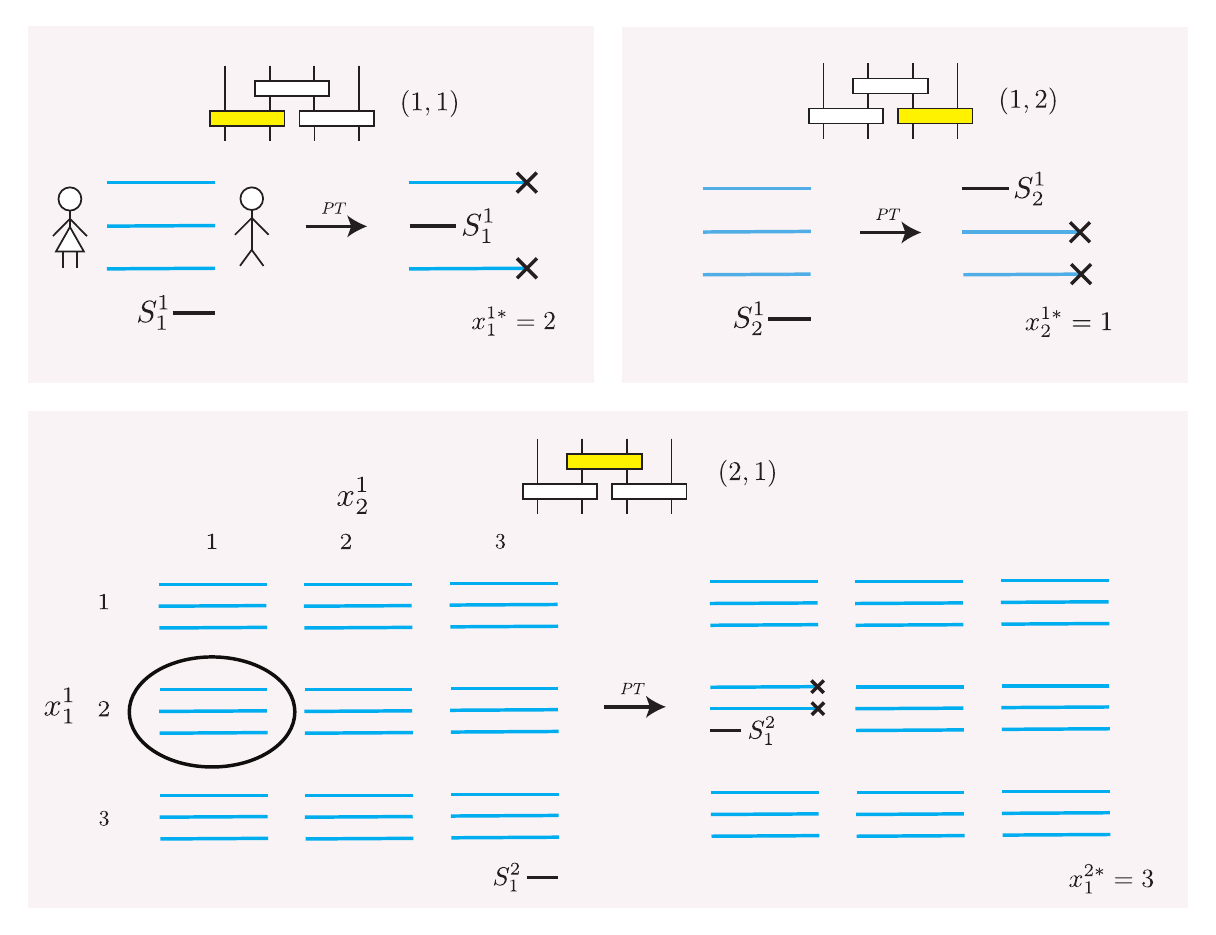}
    \caption{Port-based teleportation step of the protocol for various gates.
    The gate under consideration is highlighted in yellow, and the arrows labelled PT denote the result of port-based teleportation.
     Top left: only the resources for performing one port-based teleportation are needed for gates in the first layer.
     Bob port-teleports the two qubits the gate acts on, $S^1_1$, and learns that they have arrived on Alice's end at port number $x^{1*}_1=2$.
     Top right: Bob port-teleports the two qubits in $S^1_2$ and learns they arrive at port $x^{1*}_2=1$.
     Bottom: the systems on which this gate acts have already been port-teleported before, and Alice needs to know the value of the correct port for each of the previous teleportations in order to know which Pauli keys to apply.
     For this reason they set up resources for every possible combination of values.
     Bob then port-teleports the qubits $S^2_1$ through the resource labeled by $x^{1}_1=x^{1*}_1=1$ and $x^{1}_2=x^{1*}_2=1$.}  
    \label{fig:port-teleportation}
\end{figure*}

Now we have the notation needed to formally state the general protocol.
In this protocol, steps 2-5 are looped over many times -- once per gate in the circuit structure.
Some parts of these steps are only relevant on iterations corresponding to gates beyond the first layer of the circuit, and are reliant on later steps that occurred in previous iterations.
To aid the reader in such cases, we separately express the simpler form of those steps in the first layer.
\begin{protocol}\ \label{protocol}

\noindent\textbf{Setup:} $N$ is fixed to be a constant that controls the accuracy of the protocol. 
We also have a fixed circuit structure $\{S_j^i\}$ corresponding to a circuit decomposition $\{U_j^{i}\}$ of the unitary to be implemented, $U$.
For each $(i,j)$, Alice and Bob allocate a set of $2N^{|\fm(i,j)|}$ maximally entangled pairs of $2^{k_{i,j}}$-dimensional systems, labelled by $(i,j,x_{\fm(i,j)},m)$.
Here the notation $x_{\fm(i,j)}$ denotes a set of labels $x^{i'}_{j'} \in [N]$ for each $(i',j')\in \fm(i,j)$, and we use similar notation throughout the protocol.
The label $m\in\{A,B\}$ specifies whether the system is used for port-based teleportation from Bob to Alice $(B)$ or for normal teleportation from Alice to Bob $(A)$.
We refer to Alice's share of an entangled pair $(i,j,x_{\fm(i,j)},m)$ as $A(i,j,x_{\fm(i,j)},m)$, and similarly Bob's share as $B(i,j,x_{\fm(i,j)},m)$.
A further $n/2$ qubits are allocated for an initial teleportation from Alice to Bob.
The protocol then proceeds as follows.
\begin{enumerate}[topsep=0pt,itemsep=0ex,partopsep=1ex,parsep=1ex]
    \item Alice normal-teleports her $n/2$ qubits to Bob, resulting in him holding state $P\ket\psi$ with $P$ a Pauli known by Alice.
    It remains to implement $UP^\dagger$.
    Begin by setting $i=j=1$.
    \item \textbf{First layer:}
    Bob port-teleports $S_j^1$ to Alice through the $N$ systems labelled $B(1,j,x_j^1, B)$, from which he gets a key $x_j^{1*}\in [N]$.\\
    \textbf{Subsequent layers:}
    Bob port-teleports $S_j^i$ to Alice through the $N$ systems labelled $B(i,j,(x_j^i,x^*_{ \anc(i,j)}), B)$\footnote{This is $N$ ports since only $x_j^i$ varies; the label for each ancestor is fixed by the keys of previous port-based teleportations.}, from which he gets a key $x_j^{i*}\in [N]$. This is depicted in \cref{fig:port-teleportation}.
    \item \textbf{First layer:}
     To each of the $N$ systems labeled $A(1,j,x_j^i,B)$, Alice applies $U^1_j P^\dagger|_{S_j^1} $\footnote{For a Pauli string $P'$, $P'|_{S^i_j}$ means the string obtained by keeping only the part of the string supported on  ${S^i_j}$.}.\\
    \textbf{Subsequent layers:}
     To each of the $N^{|\fm(i,j)|}$ systems labeled $A(i,j,x_{\fm(i,j)},B)$, Alice applies $U^i_j \displaystyle{\prod_{(i',j')\in \pr(i,j)} P^\dagger_{i',j'}(x_{\fm(i',j')})|_{S^i_j}}$\footnote{Note that the notation $P^\dagger_{i',j'}(x_{\fm(i',j')})$ is defined in step 4; it is not used until the second layer.};
    that is, she looks up the normal-teleportation key corresponding to the values of the variables in $x_{\fm(i',j')} \subset x_{\fm(i,j)}$ to decrypt each port.
    \item Alice normal-teleports each system back to Bob using the entangled pair labelled $A(i,j,x_{\fm(i,j)}, A)$, receiving a separate key for each port. She labels these keys according to the corresponding port label, $x_{\fm(i,j)}$.
    She then has a lookup function $P_{i,j}(x_{\fm(i,j)})$ that gives her the appropriate Pauli to be used in subsequent instances of step 3.
    \item Bob discards all the systems $B(i,j,x_{\fm(i,j)},A)$ not labelled $x^*_{\fm(i,j)}$. 
    Repeat from step 2 for all $j$ from $1$ to $n_i$, and then for all $i$ from $1$ to $d$.
    \item \emph{Communication Round:} Bob sends Alice's $n/2$ qubits back to her, and Alice and Bob send all keys to one another.
    \item Both players fully decrypt their systems to approximately obtain the state $U\ket{\psi}_{AB}$.
\end{enumerate}
\end{protocol}

In \cref{appendix: port-error-bounds} we show that the diamond norm between the channel $\mathcal{N}$ implemented by this protocol and the desired unitary channel $\mathcal{U}$ is bounded as
\begin{align}
    \| \mathcal{N} - \mathcal{U} \|_\diamond \leq \epsilon,
\end{align}
so long as we choose
\begin{equation}\label{eq: good-value-for-N}
    N\geq  \left(\frac{4}{\epsilon}\sum_{i,j} 2^{2k_{i,j}}\right)^2.
\end{equation}

\subsection{Entanglement cost}\label{subsec: entanglement-cost}
Let us determine the entanglement cost of this protocol, ignoring the $\frac{n}{2}$ term from the initial normal teleportation from Alice to Bob, which is subleading in all regimes of interest.
We need $2N^{|\fm(i,j)|}$ sets of maximally entangled pairs of $k_{i,j}$-qubit systems for each combination of $i$ and $j$.
Thus, our total entanglement cost as measured in Bell pairs is
\begin{align}
    E &=2 \sum_{i,j} k_{i,j} N^{|\fm(i,j)|}.
\end{align}
We emphasize that this depends only on the circuit structure $\{S_j^i\}$ and not any other properties of the unitary.

For simplicity we henceforth assume that $k$ is constant.
We also define the maximal past light cone volume as $V \defi k \max_{i,j} |\fm(i,j)|$.
Using $V$ as an upper bound for $|\fm(i,j)|$ and substituting the expression for $N$ in \cref{eq: good-value-for-N} gives
\begin{align} \label{eq:cost}
     E&\leq 2 nd \left( \frac{4nd}{\epsilon k} 2^{2k}\right)^{\frac{2V}{k}}.
\end{align}
This is essentially dominated by the scaling
\begin{align} \label{eq:roughcost}
     E&\lesssim O\left( n^{4V} \right),
\end{align}
where we treat $\epsilon$ as a constant and ignore sub-leading dependencies in each variable.
Here we have also made use of $V\geq d$ and used $n 2^{2k} \leq n^{2k}$ which is always true for large $n$ regardless of how $k$ might scale with $n$.
These inequalities only mildly weaken the bound in \cref{eq:roughcost}.

Firstly, if the circuit has generic all-to-all interactions with $d$ layers of $k$-local gates, then $V = \sum_{i=1}^d k^i \sim k^d$, and so $E$ is bounded by
\begin{align} \label{eq:nlcost}
	 E &\sim O\qty(n ^{4k^d}).
\end{align}
In the case that the circuit has a geometrically local structure in $D$-dimensional Euclidean space, then $V= \alpha k d^{D+1}$ for some constant $\alpha$, and so
\begin{align}
    E&\sim O\qty( n^{4 \alpha k d^{D+1}} ). \label{eq:geocost}
\end{align}

\subsection{Regime of efficiency}\label{subsec: regime-of-efficiency}
From the expression \cref{eq:geocost,eq:nlcost}, we can make the following conclusions.
A unitary $U$ can be computed non-locally with polynomial entanglement if it can be implemented by a circuit with $k\sim O(1)$ and $d\sim O(1)$.
Furthermore, it can be computed with \emph{quasi-polynomial} entanglement if any of the following hold:
 \begin{enumerate}
     \item $k\sim \polylog n$, $d\sim O(1)$,
     \item $k\sim O(1)$, $d\sim \log \log n$, or
     \item $k\sim \polylog n$, $d\sim \polylog n$, and the circuit has geometrically local structure in $D$ spatial dimensions.
 \end{enumerate}
Without geometric locality but with $k\sim \polylog n$ and $d\sim \log \log n$, the entanglement scales as $\exp\exp\polylog\log n$, which can be thought of as ``quasi-quasi-polynomial'', and is still much slower than exponential growth.

Consider the classes described by cases 1 and 3 above.
To the authors' knowledge, there has been essentially no work studying circuits with such ``medium'' sized gates, and thus new methods are needed to characterize these classes of unitaries.
They are both strictly contained in those unitaries implementable by quasi-polynomial depth circuits of $O(1)$-local gates.
This follows by appealing to Ref.~\cite{exact-two-qubit-synthesis}, which shows that a $k$-qubit gate can be decomposed into $ 4^k$ layers of $2$-qubit gates.

One can also use non-rigorous counting arguments of the sort found in Ref.~\cite{brown-susskind} to characterize the size of such a class.
These arguments characterize the ``number of unitaries'' in a set as the ratio of the volume of the set according to some metric to the volume of a ball of radius $\epsilon$.
Including combinatorial factors, one can show that the log-number of distinct circuits\footnote{Note that the number of distinct circuits is larger than the number of distinct unitaries implemented by such a circuit; however for $d$ less than exponential in $n$ these numbers are believed to be similar \cite{brown-susskind}.}
with $d$ layers of $k$-qubit gates on $n$ total qubits scales roughly as
\begin{align}
    \log|\crc(n,k,d)|
    \sim n d\left[ \log(n) + 4^k \right] .
\end{align}
This indicates that, for example, the asymptotic scaling of the log number of unitaries implementable in class 1 above is comparable to that of the unitaries implementable by quasi-polynomial-depth circuits of $2$-qubit gates.

\subsection{Distinction from other protocols}
We will now show that the class of unitaries that our protocol can efficiently implement contains unitaries not achievable by other protocols. 
As we have already seen, the protocol of Ref.~\cite{beigi-koenig} requires entanglement exponential in the number of qubits regardless of the unitary.
Thus our protocol outperforms this one for circuits with sufficiently small light cones.

The other protocol for general unitaries that outperforms the one from Ref.~\cite{beigi-koenig} in some regimes is that of Ref.~\cite{T-gate-protocol}.
There they provide two protocols whose entanglement cost is determined by Clifford + T circuit decompositions.
One protocol scales as $n 2^{k_t}$ where $k_t$ is the number of $T$ gates in the circuit. The other scales as $(68n)^{d_t}$ where $d_t$ is the number of circuit layers containing a $T$ gate, i.e. the $T$-depth. We will show that our protocol can efficiently implement a unitary with $k_t$ and $d_t$ polynomial in $n$, a regime where the aforementioned protocols perform poorly.

The simplest example of such a unitary is one that factorizes into unitaries acting on $O(\log n)$ qubits, with each of the smaller unitaries having near-maximal $T$-depth.
This is exponential in the number of qubits acted on \cite{exact-two-qubit-synthesis}, so $k_t\sim d_t\sim \poly (n)$.
If we ran our protocol on the total unitary, treating each smaller unitary as a $k\sim \log n$-qubit gate, then the depth will be $1$ and our protocol costs $\sim n^{\log n}$, a quasi-polynomial.

Note however that in this example our protocol does nothing but perform the protocol of Ref.~\cite{beigi-koenig} many times in parallel. To obtain a less trivial example, add a second layer of $\log n$ sized gates which act across the supports of the original small unitaries. Choosing the gates of the second layer to have $O(1)$ $T$-count, we can be assured that the total unitary remains in the inefficient regime of the protocols in \cite{T-gate-protocol}. For this new unitary our protocol still requires only a quasi-polynomial amount of entanglement.

\section{Enhancements} \label{sec:extensions}

There are several complementary actions Alice and Bob can take in order to re-express the unitary they perform in a less costly manner. Firstly, they can apply local operations on their systems before and after the protocol. Secondly, they can apply arbitrary Cliffords at the beginning and end of the circuit. Finally, they can add auxiliary systems. We show how each of these is done, and give examples where the latter is useful.

\subsection{Local pre/post-processing}\label{subsec:}

Let Alice and Bob re-express $U$ as $(W_A'\otimes W_B')U_{AB}(W_A\otimes W_B)$ for some $n/2$-qubit unitaries $W_A,W_B,W_A',$ and $W_B'$ and $n$-qubit unitary $U_{AB}$. They act with $W_A$ and $W_B$ on the systems they respectively hold at the beginning of the protocol, and with $W_A'$ and $W_B'$ at the end of it. By running the protocol on $U_{AB}$ they then obtain the same result as though they had run it on $U$ without pre/post-processing. 
Crucially, the entanglement cost will depend on the circuit structure of $U_{AB}$, which may be more conducive to our protocol than that of $U$.

\subsection{Clifford sandwich}\label{subsec: clifford-sandwhich}

Suppose that $U$ has a circuit decomposition $C_fU'C_i$ with $C_i,C_f$ being $n$-qubit Cliffords and $U'$ an $n$-qubit unitary that has a circuit decomposition with structure $\{S^i_j\}$. Then we can construct an NLQC protocol that accomplishes $U$ with the same cost as our original protocol takes to accomplish any unitary via a circuit structure $\{S^i_j\}$. To do so we run the protocol as though to implement $U'$ via this circuit structure, but modified as follows.

Between steps 2 and 3, i.e. right after Alice has teleported her $\frac{n}{2}$ qubits to Bob, Bob applies the Clifford $C_i$ to his combined system. This means he now holds the state $QC_i\ket{\psi}$ where $Q=C_iPC_i^\dagger$. Note that $Q$ is a Pauli, and that Alice knows its identity since she knows both $C_i$ and $P$. If Alice and Bob then run the remainder of the original protocol but replacing $P$ with $Q$, they will have implemented $U'C_i$.

To additionally implement $C_f$, the following modification is added right before step 6, the communication round.
At this point Bob holds the state
\begin{align}
    \mathcal{P}(\vec{x}^*)U'C_i\ket{\psi}_{AB},
\end{align}
where $\mathcal{P}$ is a Pauli valued function of $\vec{x}$, all indices of the port teleportations Bob has performed throughout the protocol. Alice posses knowledge of $\mathcal{P}$ but not $\vec{x}^*$, and vice versa for Bob. Nonetheless, Bob can apply $C_f$ to the state, and the communication round can commence as in the original protocol. They then both possess knowledge of $\mathcal{P}(\vec{x}^*)$, and can jointly apply $C_f\mathcal{P}(\vec{x}^*)C_f^\dagger$, since this is a Pauli and thus factorizes between the $A$ and $B$ systems. This completes the protocol.

Local pre/post-processing can straightforwardly be performed before and after the Clifford gates. 

\subsection{Auxiliary systems}

Alice and Bob can make use of Auxiliary systems by generalizing the pre/post-processing step to make use of isometries rather than unitaries.
More precisely, suppose $U$ can be written as 
\begin{align}\label{eq: auxiliary-implementation}
    U=  (\mathcal{W}_{A\rightarrow \tilde{A}}\otimes \mathcal{W}_{B\rightarrow \tilde{B}})U_{\tilde{A}\tilde{B}}( \mathcal{V}_{A\rightarrow \tilde{A}}\otimes \mathcal{V}_{B\rightarrow \tilde{B}})
\end{align}
where $\mathcal{V}_{A\rightarrow \tilde{A}}$ and $ \mathcal{V}_{B\rightarrow \tilde{B}}$ are isometries, $\mathcal{W}_{A\rightarrow \tilde{A}}$ and $ \mathcal{W}_{B\rightarrow \tilde{B}}$ are channels, and $U_{\tilde{A}\tilde{B}}$ is a unitary.
Then the isometries and channels can be absorbed into local pre- and post-processing, and the NLQC protocol performed on $\tilde{U}$.
The entanglement cost will depend only on the circuit structure needed to implement $U_{\tilde{A}\tilde{B}}$.
Provided the number of auxiliary systems added is polynomial in $n$, and $d$ and $k$ are unchanged, the regime of efficiency will remain the same.

When would using auxiliary systems be useful? We will consider two examples: locality-preserving unitaries, and the embedding of the protocol in Ref.~\cite{code-routing}.

It is well known that locality-preserving unitaries do not always admit a circuit decomposition that reflects their causal structure \cite{unitarity+causality=localizability}.
For example, the unitary that shifts all $n$ qubits in a closed 1-d chain by one site in the same direction requires a circuit whose depth grows proportionally to the number of qubits.
However, using the techniques of Ref.~\cite{unitarity+causality=localizability}, $n$ auxiliary qubits can be introduced in the manner of \cref{eq: auxiliary-implementation} to implement this unitary with a circuit of depth $2$ and gate size $2$, which our protocol can do efficiently.
More generally these techniques can be used to decompose any locality-preserving unitary.

As a second example, we will consider the protocol introduced in \cite{code-routing} for the task of $f-$routing, a subclass of NLQC in which Alice receives a qubit and a classical string $x$, Bob receives a classical string $y$, and their task is to return the qubit to Alice or Bob depending on the value of a Boolean function $f(x,y)$, subject to the same constraints as general NLQC.
Their protocol can be written in the form of \cref{eq: auxiliary-implementation} with $\{\tilde{S^i_j}\}$ having $O(1)$ depth and $O(1)$-sized gates, and thus can be embedded into our protocol while still performed efficiently.
In their protocol, auxiliary systems are introduced when Alice's input qubit is encoded into a secret sharing scheme.
Each of these is routed to Alice or Bob depending on the value of another Boolean function that depends only on an $O(1)$ subset of the classical bits in $x$ and $y$, and so can be implemented as an $O(1)$-sized gate.
These gates can be parallelized by making as many copies of $x$ and $y$ in the pre-processing stage as needed for the cases when the routing of several different shares depends on the same bits.    
In particular, only a polynomial number of additional copies are needed because the construction of Ref.~\cite{code-routing} requires that the number of gates be polynomial in $n$.

\section{Conclusion}
As with NLQC in general, our results are relevant for both theory and practice.
On the practical side, our result can be considered as a new class of attacks on position-based cryptography, which is based on the premise that an attacker does not have access to a sufficient amount of entanglement in order to perform a computation. If the computation is in the class we have described, however, the amount required is quasi-polynomial in the system size and so may not be large enough to be prohibitive. 

Our results may also be of interest for modular quantum computing architectures (see e.g.\ \cite{modular}), which are composed of several parallelized computational components capable of performing computation on some fixed number of qubits with constraints on the communication between them (e.g.\ due to spatial separation).
Our results show that such architectures can at least perform computations in the class we have described with only one round of communication.

On the theoretical side, our results are in line with a prediction made about quantum information from studying holography, that polynomial-depth circuits should also be implementable with efficient entanglement cost.
This further establishes the deep connection between holography and NLQC.
In return, understanding how NLQC can be performed efficiently may give insight into how dynamics occur in holography, which have been notoriously difficult to capture in discrete models.

Future work is needed to better understand the classes of efficiently computable unitaries described in \cref{subsec: regime-of-efficiency}, and how they relate to more familiar complexity classes of unitaries.
In particular, it remains to be found which, if any, polynomial-depth circuits of two-qubit gates are outside the regime of efficiency of our protocol when accounting for the extensions of \cref{sec:extensions}.
Finally, it may be possible to find a better method to deal with the issue of the port dependence of Pauli keys, as this is the greatest bottleneck of our
protocol.

\section*{Acknowledgement}
We thank Alex May for suggesting the approach of exploiting a unitary's decomposition into a circuit of medium-sized gates to make more efficient use of port teleportation.
We also thank Nick Hunter-Jones, Adam Brown and Adam Bouland for useful discussions.
SC is supported by a graduate fellowship award from Knight-Hennessy Scholars at Stanford University.

\appendix

\section{Protocol error bounds} \label{appendix: port-error-bounds}

Let $\mathcal{N}$ be the channel applied to an input state of the protocol in \ref{sec: protocol} for a unitary $U$ decomposed into $\{U^i_j\}$ with circuit structure $\{S_i^j\}$, with gate size $|S_i^j|=k_{i,j}$. Let $\mathcal{U}:=U\cdot U^\dagger$.
We would like to determine an upper bound on $||\mathcal{N}-\mathcal{U}||_\diamond$.

First we need to write down the explicit form of $\mathcal{N}$. To do so define the following quantities
\begin{itemize}
    \item $\mathcal{E}^i_j$ is the error channel defined by port-teleporting $S_i^j$ using $N$ maximally entangled $k_{i,j}$ qubit states, and keeping the correct port.
    \item $\mathcal{E}_i:= \mathcal{E}^i_1\cdots \mathcal{E}^i_{n_i}$
    \item $\mathcal{U}_i:= U^i\cdot U^{i\dagger}$ where $U^i=U^i_1\cdots U^i_{n_i}$. Thus $\mathcal{U}=\mathcal{U}_d\cdots\mathcal{U}_1$.
    \item $\mathcal{P}_i:= P^i\cdot P^{i\dagger}$ where\\ $P^i=\prod_{(i',j')\in \pr(i,j)} P_{i',j'}(x^*_{\fm(i',j')})|_{S^i_j}$ for $i\geq 1$ and $P^0=P$.
    \item $\mathcal{E}'_i= \mathcal{P}_{i-1}^\dagger\mathcal{E}_i\mathcal{P}_{i-1}$
\end{itemize}

Then we can write $\mathcal{N}$ as 
\begin{align*}
    \mathcal{N}=\mathcal{U}_d\mathcal{E}'_d\cdots
    \mathcal{U}_1\mathcal{E}'_1.
\end{align*}

We will need to make use of the following properties of the diamond norm (see section 3.3 of Watrous' book where it is referred to as the "completely bounded trace norm").

Let $\Phi_0$,$\Phi_1$,$\Psi_0$,$\Psi_1$ be CPTP maps from $\mathcal{B}(A)$ to $\mathcal{B}(A)$ for some quantum system $A$. Then
\begin{align} \label{eq: convenient-watrous-equation}
    &||\Psi_1\Psi_0-\Phi_1\Phi_0||_\diamond\leq \nonumber\\ &||\Psi_0-\Phi_0||_\diamond + ||\Psi_1-\Phi_1||_\diamond
\end{align}

\begin{align} \label{eq: diamond-of-channel}
&||\Phi_0||_\diamond = 1
\end{align}

If $\Phi_0$, and $\Phi_1$ are instead general linear maps then we have
\begin{align}\label{eq: diamond-norm-factor}
&||\Phi_0\otimes \Phi_1||_\diamond = ||\Phi_0||_\diamond ||\Phi_1||_\diamond
\end{align}

We can use equation \ref{eq: convenient-watrous-equation} to get
\begin{align*}
    ||\mathcal{N}-\mathcal{U}||_\diamond 
    &\leq \sum_{i=1}^d ||\mathcal{U}_i\mathcal{E}'_i-\mathcal{U}_i||_\diamond\\
    &\leq\sum_{i=1}^d ||\mathcal{E}'_i-\mathcal{I}_{\mathbb{C}^{2^n}}||_\diamond\\
    &\leq\sum_{i=1}^d ||\mathcal{P}_{i-1}^\dagger\mathcal{E}_i\mathcal{P}_{i-1}-
    \mathcal{P}_{i-1}^\dagger\mathcal{P}_{i-1}||_\diamond\\
    &\leq\sum_{i=1}^d ||\mathcal{P}_{i-1}^\dagger\mathcal{E}_i-
    \mathcal{P}_{i-1}^\dagger||_\diamond\\
    &\leq\sum_{i=1}^d ||\mathcal{E}_i-\mathcal{I}_{\mathbb{C}^{2^n}}||_\diamond\\
    &\leq d ||\mathcal{E}_i-\mathcal{I}_{\mathbb{C}^{2^n}}||_\diamond.
\end{align*}
where in the last line we used that the argument of the sum does not depend on $i$. Writing $\mathcal{E}_i:= \mathcal{E}^i_1\cdots \mathcal{E}^i_{n_i}$ and again using equation \ref{eq: convenient-watrous-equation} we get
\begin{align*}
    ||\mathcal{N}-\mathcal{U}||_\diamond \leq \sum_{i,j} ||\mathcal{E}^i_j-\mathcal{I}_{\mathbb{C}^{2^n}}||_\diamond.
\end{align*}
Finally, use the fact that $\mathcal{E}^i_j=\mathcal{E}^i_j|_{S^i_j}\otimes\mathcal{I}_{\mathbb{C}^{2^{n-k_{i,j}}}}$ and equations \ref{eq: diamond-norm-factor}, and \ref{eq: diamond-of-channel} to get
\begin{align*}
    ||\mathcal{N}-\mathcal{U}||_\diamond \leq \sum_{i,j} ||\mathcal{E}^i_j|_{S^i_j}-\mathcal{I}_{\mathbb{C}^{2^{k_{i,j}}}}||_\diamond.
\end{align*}
From \cite{beigi-koenig} we have that
\begin{align*}
    ||\mathcal{E}^i_j|_{S^i_j}-\mathcal{I}_{\mathbb{C}^{2^k}}||_\diamond\leq 4\frac{2^{2k}}{\sqrt{N}}.
\end{align*}
Thus the bound we desire is 
\begin{align*}
    ||\mathcal{N}-\mathcal{U}||_\diamond \leq 4 \sum_{i,j} \frac{2^{2k_{i,j}}}{\sqrt{N}}.
\end{align*}
If we want the norm to be less than $\epsilon$, then we can choose
\begin{align} 
    N= \frac{16}{\epsilon^2} \left(\sum_{i,j} 2^{2k_{i,j}}\right)^2.
\end{align}

\bibliographystyle{unsrtnat}
\bibliography{biblio}

\end{document}